\documentstyle{l-aa} 

\input epsf.tex 
\newcommand{\mincir}{\raise -2.truept\hbox{\rlap{\hbox{$\sim$}}\raise5.truept 
\hbox{$?$}\ }} 
\newcommand{\gr}{\kern 2pt\hbox{}^\circ{\kern -2pt K}} 
\newcommand{\magcir}{\raise -2.truept\hbox{\rlap{\hbox{$\sim$}}\raise5.truept 
\hbox{$?$}\ }} 
 
\newcommand{\Om}{\Omega} 
\newcommand{\apj}{ApJ}
\newcommand{\apjl}{ApJL}
\newcommand{\mnras}{MNRAS}
\newcommand{\aap}{AA}

\begin{document} 

\thesaurus{
       (12.03.4;  
        12.04.1;  
        12.12.1;  
Universe
        11.03.1;  
          }
 
\title{ An analytic approximation of MDM power spectra 
in four dimensional parameter space}

\author{B.~Novosyadlyj \inst{1}, R.~Durrer \inst{2}, V.N.~Lukash \inst{3} } 

\offprints{Bohdan Novosyadlyj}
 
\institute{Astronomical Observatory of L'viv State University, Kyryla and 
Mephodia str.8, 290005, L'viv, Ukraine
\and Department of Theoretical Physics, University of Geneva, 
Ernest Ansermet, CH-1211 Geneva 4, Switzeland
\and Astro Space Center of Lebedev Physical Institute of RAS, 
Profsoyuznaya 84/32, 117810 Moscow, Russia}  
 
\date{Received \dots; accepted \dots}
\maketitle
\markboth{An analytic approximation for MDM spectra}{}

\begin{abstract} 
An accurate analytic approximation of the transfer function for the 
power spectra of primordial density perturbations in mixed dark matter
models is presented. The fitting formula in a matter-dominated
 Universe ($\Omega_0=\Omega_M=1$) is a function of
wavenumber $k$, redshift $z$ and four cosmological  parameters: the  
density of massive neutrinos, $\Omega_{\nu}$, the number  of massive 
neutrino species, $N_{\nu}$, the baryon density, $\Omega_{b}$ and the 
dimensionless Hubble constant, $h$. Our formula is accurate in a broad 
range of parameters: 
$k\le 100\;h/Mpc$, $z\le 30$, $\Omega_{\nu}\le 0.5$, $N_{\nu}\le 3$, 
$\Omega_{b}\le 0.3$, $0.3\le h\le 0.7$. The deviation of the variance 
of density fluctuations calculated with our formula from numerical results 
obtained with CMBfast is less than $6\%$ for the entire range of parameters.
It increases with $\Omega_bh^{2}$  and is less than $\le 3\%$
for $\Omega_bh^{2}\le 0.05$.

The performance of the analytic approximation of MDM power spectra
proposed here  is compared with other approximations found in the
literature (\cite{hol89,pog95,ma96,eh3}). Our approximation turns out
to be  closest to numerical results in the parameter space considered here.
 
\end{abstract} 
 
 
\section{Introduction} 
 
Finding a viable  model for the formation of large scale structure
(LSS) is an important problem in cosmology.
 Models with a minimal number of free parameters, such as standard 
cold dark matter (sCDM) or standard cold plus hot, mixed 
dark matter (sMDM) only marginally match observational 
data. Better agreement between  predictions and observational 
data can be achieved in  models with a larger numbers of parameters 
(CDM or MDM with baryons, tilt of primordial power spectrum,
3D curvature, cosmological constant, see, {\it e.g.}, 
\cite{vkn98} and refs. therein). 
In view of the growing amount of observational data, we seriously 
have to discuss the precise quantitative differences
between theory and observations for the whole class of
available models by varying all the input parameters such as 
the tilt of primordial spectrum, $n$, the density of cold dark matter, 
$\Omega_{CDM}$, hot dark matter, 
$\Omega_{\nu}$, and  baryons, $\Omega_b$, the vacuum energy or
cosmological constant, $\Omega_{\Lambda}$, and the 
Hubble parameter $h$ ($h=H_0/100\;km/s/Mpc$), to find the values which
agree best with observations of large scale structure
(or even to exclude the whole family of models.).

Publicly available fast codes to calculate the  transfer function 
and power spectrum of fluctuations in the cosmic microwave background 
(CMB) (\cite{sz96}, CMBfast) are an essential ingredient in this process.
But even CMBfast is too bulky and too slow for an effective search 
of cosmological parameters by means of a  $\chi^2$-minimization, like that 
of Marquardt (see \cite{nr92}). To solve this problem, analytic 
approximations of the transfer function are of great value. 
Recently, such an approximation has been proposed by \cite{eh3}
(this reference is denoted by EH2 in the sequel). 
Previously, approximations
by \cite{hol89,pog95,ma96} have been used. 
 
 Holtzman's 
approximation is very accurate but it is an approximation for fixed
cosmological parameters. Therefore it can not be useful 
for the purpose mentioned above. The analytic approximation by
\cite{pog95} is valid in the 2-dimensional parameter space $(\Omega_{\nu}, 
h)$, and  $z$ (the redshift). It has the correct asymptotic behavior at small 
and  large $k$, but the systematic error of the transfer function $T(k)$ 
is relatively large (10\%-15\%) in the 
 important range of scales  $0.3\le k\le 10\;h/$Mpc. This error,
however introduces discrepancies of 4\% to 10\% in $\sigma_R$ which
represents an integral over $k$.
 Ma's analytic approximation  is 
slightly more accurate in this range, but has an incorrect asymptotic behavior 
at large $k$, hence it cannot be used  for the analysis of the 
formation of small scale objects (QSO, damped $Ly_\alpha$  systems, 
$Ly_\alpha$ clouds etc.). 
 
Another weak point of these analytic 
approximations is their lack of dependence on the baryon density. 
Sugiyama's correction of the CDM transfer function in the 
presence of baryons  (\cite{bar86,sug95}) 
works well only for low baryonic content. Recent data on the high-redshift 
deuterium abundance (\cite{tyt96}), on clustering at $100$Mpc$/h$ 
 (\cite{eh4}) and 
new theoretical interpretations of the $Ly_\alpha$ forest (\cite{wei97}) 
suggest that  $\Omega_{b}$ may be higher than  the standard 
nucleosynthesis value. 
 Therefore  pure CDM and MDM models have to be modified. 
(Instead of raising  $\Omega_b$, one can also look for other
solutions, like, {\it e.g.} a cosmological constant, see below.) 
 
For CDM this has been achieved by Eisenstein $\&$ Hu (1996,
1997a\footnote{This reference is denoted by EH1 in this paper.}) using an 
analytical approach for the 
description of  small scale cosmological perturbations in the 
photon-baryon-CDM system. Their analytic 
approximation for the matter transfer function
in 2-dimensional  parameter space
($\Omega_{M}h^2$, $\Omega_b/\Omega_{M}$)
reproduces acoustic oscillations, and is quite accurate for $z<30$ (the 
residuals are smaller than 5\%) in the range $0.025\le \Omega_{M}h^{2}\le 0.25$,
$0\le \Omega_{b}/\Omega_{M}\le 0.5$, 
where $\Omega_M$ is the matter density parameter.

In EH2 an analytic approximation of the matter transfer
function  for MDM models is proposed  for a wide range of parameters
($0.06\le \Omega_{M}h^{2}\le 0.4$, $\Omega_b/\Omega_{M}\le 0.3$, 
$\Omega_{\nu}/\Omega_{M}\le 0.3$ and $z\le 30$). It is more accurate than 
previous approximations by \cite{pog95,ma96} but not as precise
as the one for the CDM+baryon model. The baryon oscillations are
mimicked  by a smooth function, therefore the approximation looses
accuracy  in the important range $0.03\le k\le 0.5~h/$Mpc. 
For the parameter choice $\Omega_{M}=1$, $\Omega_{\nu}=0.2$,
$\Omega_b=0.12$, $h=0.5$,  {\it e.g.},  the systematic residuals are about
6\% on these scales. For higher $\Omega_{\nu}$ and $\Omega_{b}$ they
become even larger.

For models with cosmological constant, the motivation to go to high
values for $\Om_\nu$ and $\Om_b$ is lost, and the parameter space
investigated in EH2 is sufficient. Models without cosmological
constant, however, tend to require relatively high baryon  or HDM
content.  In this paper, our goal is thus to construct a very 
precise analytic approximation for the redshift dependent transfer 
function in the 4-dimensional space  of spatially flat matter dominated
MDM models, $T_{MDM}(k;\Omega_{\nu},N_{\nu},\Omega_{b},h;z)$, 
which is valid for $\Omega_M =1$ and allows for 
high values of $\Om_\nu$ and $\Om_b$. In 
order to keep the baryonic features, we will use the EH1
transfer function for the cold particles+baryon system, 
$T_{CDM+b}(k;\Omega_{b},h)$, and then correct 
it for the presence of HDM by a function 
$D(k;\Omega_{\nu},N_{\nu},\Omega_{b},h;z)$, making use of the exact 
 asymptotic solutions. The resulting MDM transfer function is the 
product $T_{MDM}(k)=T_{CDM+b}(k)D(k)$. 
 
To compare our approximation with the numerical result, we use the
publicly available code 'CMBfast' by Seljak $\&$ Zaldarriaga 1996.

The paper is organized as follows: In Section 2 a short description of
the physical parameters which affect the shape of the MDM transfer 
function is given. In Section 3 we derive the analytic approximation for 
the function $D(k)$. The precision of our approximation for $T_{MDM}(k)$, 
the parameter range where it is applicable, and a comparison with the 
other results are discussed in Sections 4 and 5. In Section~6 we
present our conclusions. 
 
\section{Physical scales which determine the form of MDM transfer function} 

We assume the usual cosmological paradigm: scalar primordial density
perturbations which are generated in the early Universe, evolve in a
multicomponent medium of relativistic (photons and massless neutrinos) and
non-relativistic (baryons, massive neutrinos and CDM)
particles. Non-relativistic matter dominates the density today, 
$\Omega_{M}=\Omega_b+\Omega_{\nu}+\Omega_{CDM}$. This model is usually
called 'mixed dark matter' (MDM). The total energy density may also include a
vacuum energy, so that $\Omega_0=\Omega_M+\Omega_\Lambda$. However,
for reasons mentioned in the introduction, here we investigate 
 the case of a matter-dominated flat Universe with $\Omega_M=1$ and 
$\Omega_\Lambda=0$. Even though $\Omega_\Lambda\neq 0 $ seems to be
favored by some of the present data, our main point, allowing for high values
of $\Omega_b$, is not important in this case and the approximations by
EH2 can be used.

Models with hot dark matter or MDM have been described in the literature
by \cite{fan84,sha84,vb85,hol89,luk91}, Davis, Summers $\&$ Schlegel 1992,
\cite{sch92,van92}, 
Pogosyan $\&$ Starobinsky 1993, 1995, \cite{nov94}, Ma $\&$ Bertschinger 1994, 
1995, \cite{sz96}, EH2, \cite{vkn98} and refs. therein. Below, we simply present 
the physical parameters
which determine the shape of the MDM transfer function and which will be used 
explicitly in the  approximation which we derive here\footnote{Recall 
the definitions and relationship
between the MDM and the partial transfer functions
$$T_{MDM}=\Omega_{CDM}T_{CDM}+\Omega_\nu T_\nu +\Omega_b T_b\;,$$
$$T(k)\equiv {\delta (k,z)\over \delta (0,z)}
{\delta (0,z_{in})\over \delta (k,z_{in})}\;\;,$$ where $\delta (k,z)$
is the density perturbations in a given component and $z_{in}$ is a very high 
redshift at which all scales of interest are still super horizon.}.

\begin{figure}[th]
\epsfxsize=8truecm
\epsfbox{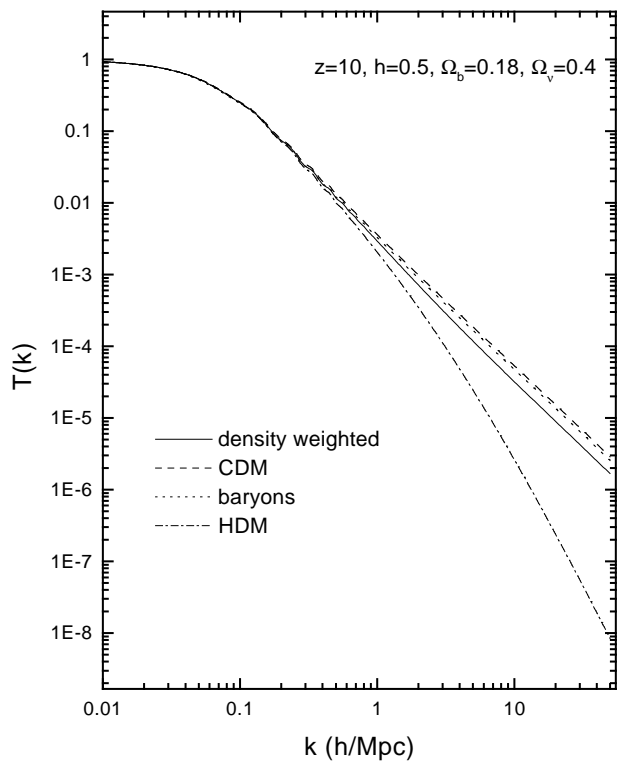}
\caption{The transfer function of density perturbations of CDM, baryons
and HDM at $z=10$ (calculated numerically).}
\end{figure}

\begin{figure}[th]
\epsfxsize=8truecm
\epsfbox{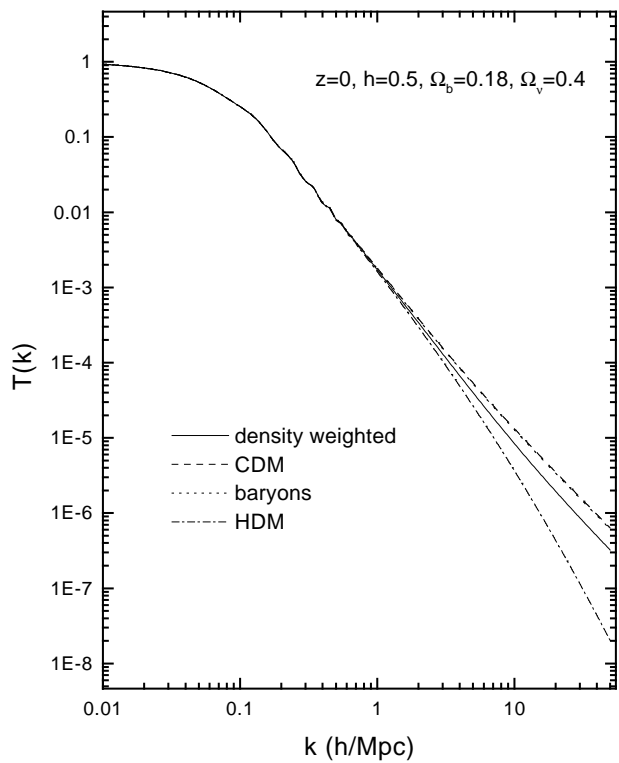}
\caption{The same as in Fig.1 but for z=0.}
\end{figure}

Since cosmological perturbations cannot grow significantly in a
radiation dominated universe, an important parameter 
is the time of equality between the densities 
of matter and radiation
$$z_{eq}=\frac{2.4\times 10^4}{1-N_{\nu}/7.4}h^{2}t_{\gamma}^{-4}-1,\eqno(1)$$
where $t_{\gamma}\equiv T_{\gamma}/2.726 K$ is the CMB temperature today,
$N_{\nu}$=1, 2 or 3 is the  number of species of  massive neutrinos
with equal mass (the number of massless neutrino species is then $3-N_{\nu}$).
The scale of the particle horizon at this epoch,
$$k_{eq}=4.7\times 10^{-4}\sqrt{1+z_{eq}}\;h/Mpc,\eqno(2)$$
is imprinted in the matter transfer function: perturbations
on smaller scales ($k>k_{eq}$) can only start growing after $z_{eq}$,
while those on larger scales ($k<k_{eq}$) keep growing at any time. 
This leads to the suppression of the transfer
function at $k>k_{eq}$. After $z_{eq}$ the fluctuations in the CDM component
are gravitationally unstable on all scales.
The scale $k_{eq}$ is thus the single physical 
parameter which determines the form of the CDM transfer function.

The transfer function for HDM ($\nu$) is more complicated because
 two more physical scales enter the problem. The time and horizon scale when 
neutrino become non-relativistic ($m_\nu\simeq 3T_\nu$) are given by
$$z_{nr}=x_{\nu}(1+z_{eq})-1\;\;,$$ 
$$k_{nr}=3.3\times 10^{-4}\sqrt{x_{\nu}(1+x_{\nu})(1+z_{eq}})\;h/Mpc,\eqno(3)$$
where $x_{\nu}\equiv \Omega_{\nu}/\Omega_{\nu\;eq}\;$, 
$\;\;\Omega_{\nu\;eq}\simeq N_\nu /(7.4-N_\nu )$ 
is the density parameter for a neutrino component becoming non-relativistic just 
at $z_{eq}$.
 The neutrino mass can be expressed in terms of $\Om_\nu$ and $N_\nu$ as
(\cite{pb93}) $m_{\nu}=94\Omega_{\nu}h^{2}N_{\nu}^{-1}t_{\gamma}^{-3}\;$eV.  
 
  The neutrino free-streaming (or Jeans\footnote{Formally the 
Jeans scale is 22.5$\%$  less than the free-streaming
scale (Bond $\&$ Szalay 1983, Davis, Summers $\&$ Schlegel 1992), 
however, $k_F$ is the relevant physical parameter
for collisionless neutrini.}) scale at $z\le z_{nr}$ is
$$k_{F}(z)\simeq 59\sqrt{{1\over 1+z_{eq}}+{1\over 1+z}}\;
\Omega_{\nu}N_{\nu}^{-1}t_{\gamma}^{-4}\;h^3/Mpc,\eqno(4)$$
which corresponds to the distance a neutrino travels in one Hubble time,
with the characteristic velocity 
$v_{\nu}\simeq {1\over x_{\nu}}{1+z\over 1+z_{eq}}.$ 
Obviously, $k_F\ge k_{nr}$, and  $k_{nr}{}^{>}_{<}k_{eq}$ for
$\Omega_{\nu}{}^{>}_{<}\Om_{\nu\;eq}\;$.

The amplitude of $\nu$-density 
perturbation on small scales ($k>k_{nr}$) is reduced in comparison
with large scales ($k<k_{nr}$). For scales larger than the free-streaming 
scale ($k<k_F$) the amplitude of density perturbations
grows in all components like $(1+z)^{-1}$ after $z_{eq}$. 
Perturbations on scales below the free-streaming scale ($k>k_F$)
are suppressed by free streaming which is imprinted in the transfer 
function of HDM. Thus the latter should be parameterized by two ratios:
$k/k_{nr}$ and $k/k_F$.

The transfer function of the baryon component is determined by the 
sound horizon and the Silk damping scale at the time of recombination 
(for details see EH1).

In reality the transfer function of each component is more complicated due to
interactions between them. At late time ($z<20$), 
the  baryonic transfer function  is practically
equal to the one of CDM, for models with $\Omega_{b}<\Omega_{CDM}$ 
 (see Figs.~1,2).
After $z_{eq}$, the free-streaming scale decreases with time (neutrino
momenta decay with the expansion of the Universe whereas the Hubble
time grows only as the square root of the scale factor, see Eq.~(4)), and
neutrino density perturbations at smaller and smaller scales become
gravitationally unstable and cluster with the CDM+baryon component. 
Today the $\nu$ free-streaming  scale may lie in the range of
galaxy to clusters scales depending on the $\nu$ mass.
On smaller scales the growing mode of perturbation is concentrated
in the CDM and baryon components. Matter density perturbation
on these scales grow like $\sim t^{\alpha}$, where 
$\alpha=(\sqrt{25-24\Omega_{\nu}}-1)/6$ (\cite{dor80}).

\section{An analytic approximation for the MDM transfer function}

To construct the MDM transfer function we use the analytic
approximation of EH1 for the transfer function of cold particles plus 
baryons and correct it for the presence of a $\nu$-component like
 \cite{pog95} and \cite{ma96}: 
$$T_{MDM}(k)=T_{CDM+b}(k)D(k)~.\eqno(5)$$ 
The function $D(k)$ must have the following asymptotics:
$$D(k\ll k_{nr})=1,$$
$$D(k\gg k_F)=(1-\Omega_{\nu})\left({1+z\over
1+z_{eq}}\right)^{\beta},$$
where $\beta=1-1.5\alpha$.
After some numerical experimentation we arrive at
 the following ansatz which satisfies these asymptotics
$$
D(k)=\left[{1+(1-\Omega_{\nu})^{1/\beta}{1+z\over 1+z_{eq}}
({k_F\over k_{nr}})^{3} \Sigma _{i=1}^{3}
\alpha_{i}\left({k\over k_F}\right)^i
\over 1+(\alpha_{4}k/k_{nr})^{3}}\right]^{\beta}.\eqno(6)
$$

We  minimize the
residuals in intermediate region ($k_{nr}<k<k_F$) by determining
$\alpha_{i}$ as best fit coefficients
by comparison with the numerical results. 

By $\chi^2$ minimization (\cite{nr92}) we first determine the
 dependence of the coefficients  $\alpha_{i}$ on $\Omega_{\nu}$
 keeping all other
parameters  fixed, to obtain an analytic approximation
$\alpha_{i}(\Omega_{\nu},z)$. The main dependence  of 
$T_{MDM}(k)$  on $\Omega_{b},\;N_{\nu},\;h$ and $z$ is taken care of by the
 dependence of $T_{CDM+b}$, $k_{nr}$, $k_F$ and of the
asymptotic solution on these parameters.  We then 
correct $\alpha_{i}$ by minimization of the residuals to include the
slight dependence on these parameters. 

Finally, the correction coefficients have the following form:
$$\alpha_{i}=a_{i}A_{i}(z)B_{i}(\Omega_b)C_{i}(h)D_{i}(N_{\nu}),$$
where $a_{i}=a_{i}(\Omega_{\nu})$,  
$A_{i}(0)=B_{i}(0.06)=C_{i}(0.5)=D_{i}(1)=1$.
The functions $A_{i}$
 depend also on $\Omega_{\nu}$.  

For all our calculations we assume a CMB  temperature of
 $T_{\gamma}=2.726K$ (\cite{mat94,kog96}).

\subsection{Dependence on $\Omega_{\nu}$ and $z$.}

We first set $h=0.5$, $\Omega_{b}=0.06$,  $N_{\nu}=1$
and determine $\alpha_{i}$ for
$\Omega_{\nu}=0.1,\;0.2,\;0.3,\;0.4,\;0.5$ and $z=0,\;10,\;20$.
We then approximate  $D(k)$ by setting  
$\alpha_{i}=a_{i}A_{i}(z)$, where
$A_i(z)=(1+z)^{b_i+c_i(1+z)}$.  The dependences
of $a_i$, $b_i$ and $c_i$ on $\Omega_{\nu}$, as well as
$B_{i}(\Omega_b)$, $C_{i}(h)$ and $D_{i}(N_{\nu})$ are given in the  Appendix.
The functions $D(k)$ for different $\Omega_{\nu}$ and its fractional 
residuals are  shown in Figs.~3 and~4.

\begin{figure}[t]
\epsfxsize=8truecm
\epsfbox{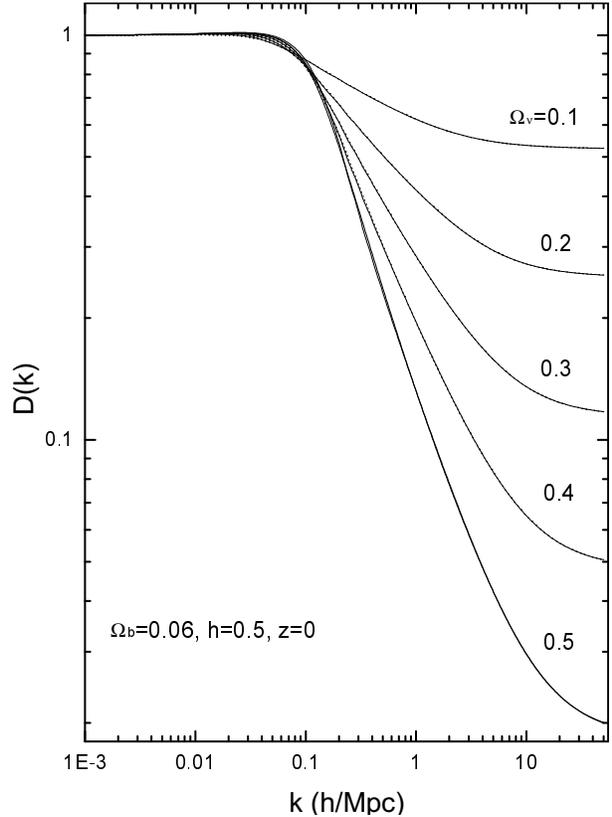}
\caption{$D(k)=T_{MDM}(k)/T_{CDM+b}(k)$
as calculated numerically (solid line) and our analytic
approximation (dotted line) for different values of $\Omega_{\nu}$. The
numerical results and the approximations overlay perfectly.}
\end{figure}

\begin{figure}[th]
\epsfxsize=8truecm
\epsfbox{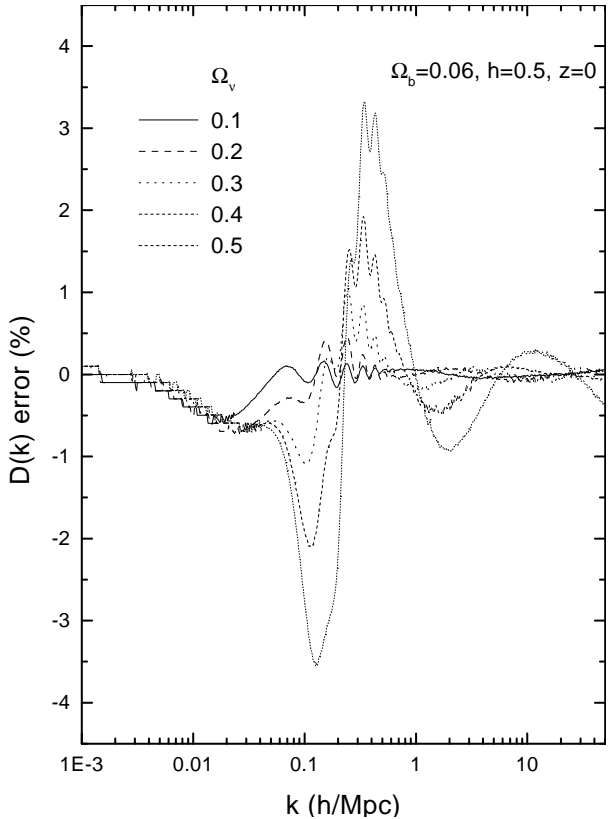}
\caption{Fractional residuals of $D(k)$ given by
$(D(k)-D_{CMBfast}(k))/D_{CMBfast}(k)$ for different values of $\Omega_{\nu}$.}
\end{figure}

We now analyze the accuracy of our analytic approximation for the MDM
transfer function $T_{MDM}(k)=T_{CDM+b}D(k)$ which in addition to 
the errors in $D(k)$ contains also those of $T_{CDM+b}(k)$ (EH1).
We define the fractional residuals for $T_{MDM}(k)$ by
$(T(k)-T_{CMBfast}(k))/T_{CMBfast}(k)$. In Fig.~5 the numerical result
for $T_{MDM}(k)$ (thick solid lines) and the
analytic approximations (dotted thin lines) are shown for different
$\Omega_{\nu}$. The fractional residuals for the latter are given in Fig.~6.
Our analytic approximation of $T_{MDM}(k)$ is sufficiently accurate for a
wide range of redshifts (see Fig.7). For $z\le 30$ the fractional residuals 
do not change by more than 2\% and stay small.

\begin{figure}[th]
\epsfxsize=8truecm
\epsfbox{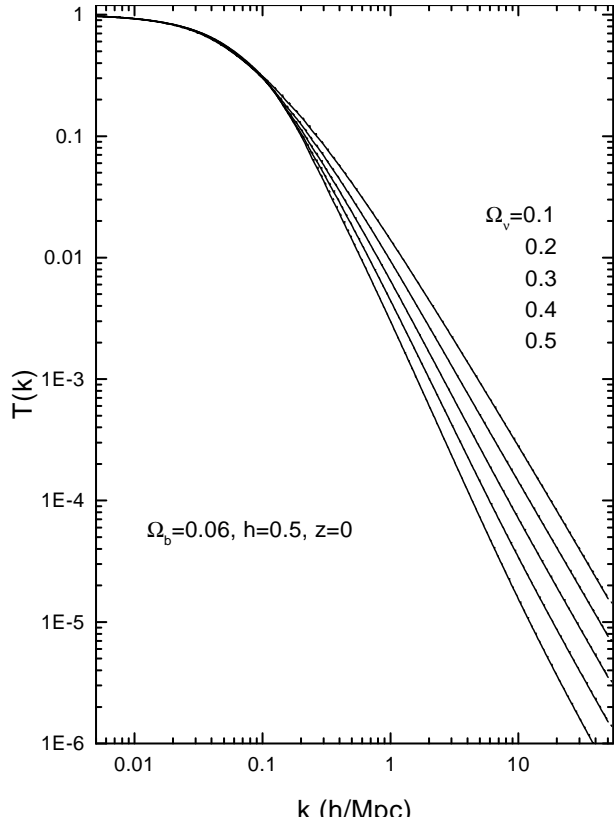}
\caption{The numerical MDM transfer function
at $z=0$  (thick solid lines)
and the analytic approximations $T_{MDM}(k)=T_{CDM+b}D(k)$ (dotted thin
lines, which perfectly overlay with the numerical result).}
\end{figure}

\begin{figure}[th]
\epsfxsize=8truecm
\epsfbox{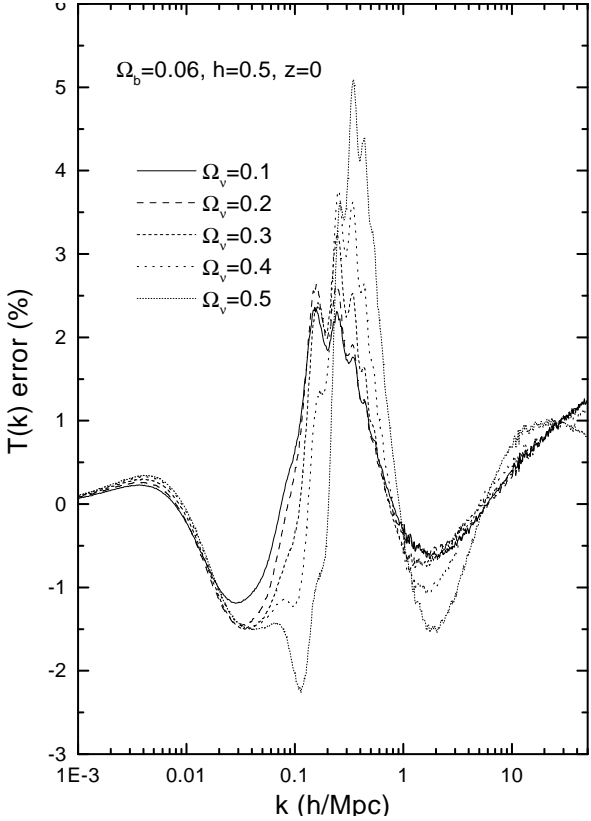}
\caption{The fractional residuals of $T(k)$ given by $(T(k) -
T_{CMBfast}(k))/T_{CMBfast}(k)$ for different values of $\Omega_{\nu}$.}.
\end{figure}

\begin{figure}[th]
\epsfxsize=8truecm
\epsfbox{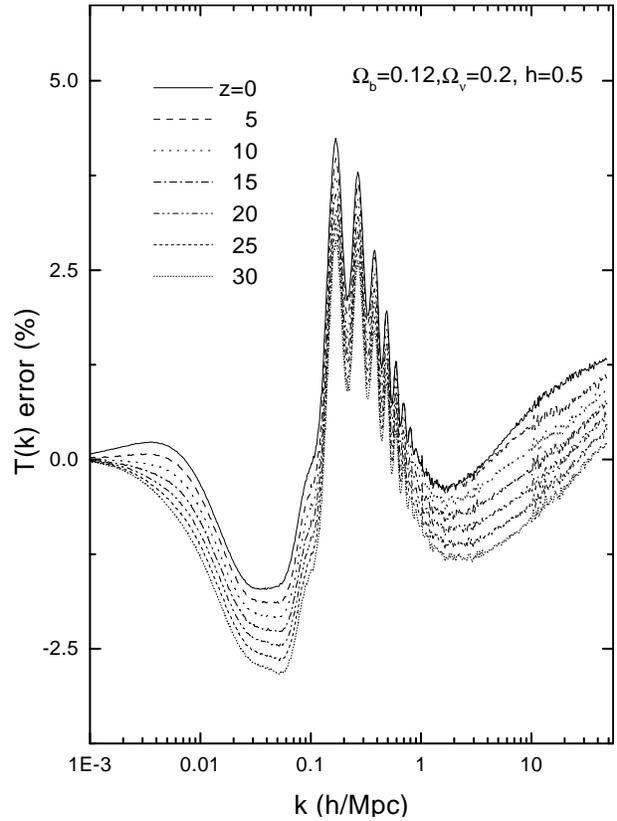}
\caption{Fractional residuals of the analytic approximation  for the MDM
transfer function at different redshifts.}
\end{figure}

\subsection{Dependence on $\Omega_{b}$ and $h$.}

\begin{figure}[th]
\epsfxsize=8truecm
\epsfbox{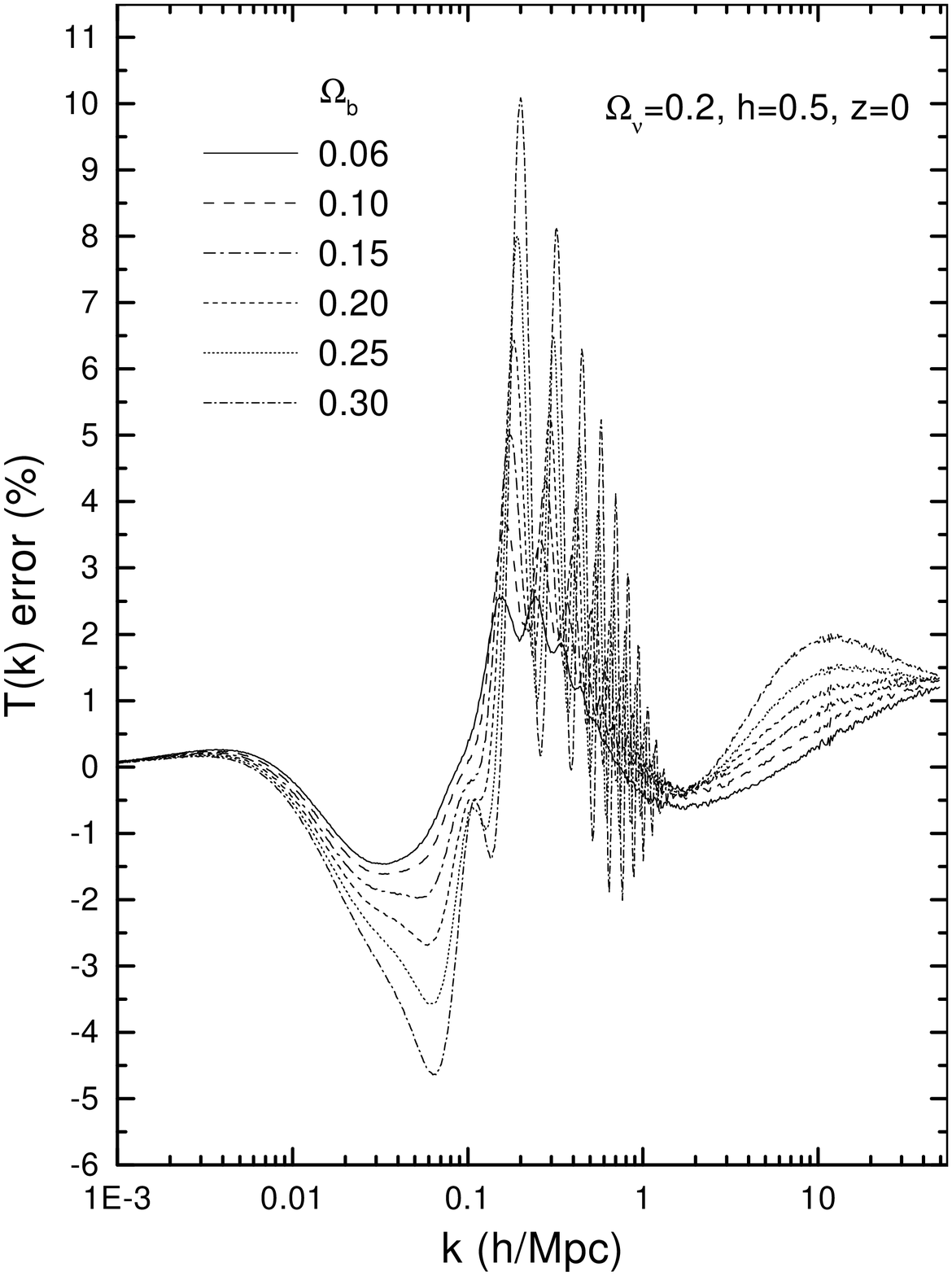}
\caption{Fractional residuals of the analytic approximation  of the MDM
transfer function for different values of $\Omega_{b}$.}
\end{figure}

We now  vary $\Omega_{b}$  fixing  different values of $\Omega_{\nu}$
and setting the other parameters $h=0.5$, $N_{\nu}=1$. We analyze
the ratio $D(k;\Omega_{\nu},\Omega_{b})/D(k;\Omega_{\nu},\Omega_{b}=0.06)$.
Since the dominant dependence of $T_{MDM}(k)$ on $\Omega_{b}$ is
already taken care of in
$T_{CDM+b}(k)$, $D(k)$ is only slightly corrected for this
parameter. Correction factors $B_i(\Omega_b)$ ($\sim 1$) as a second order
polynomial dependence on $\Omega_b/0.06$ with best-fit coefficients are
presented in the Appendix.   
The fractional residuals of
$T_{MDM}(k)$ for different $\Omega_{b}$ are shown in Fig.~8.

The maximum of the residuals grows for higher baryon fractions. 
This is due to the acoustic oscillations which become more prominent and
their analytic modeling in MDM models is more complicated.

\begin{figure}[th]
\epsfxsize=8truecm
\epsfbox{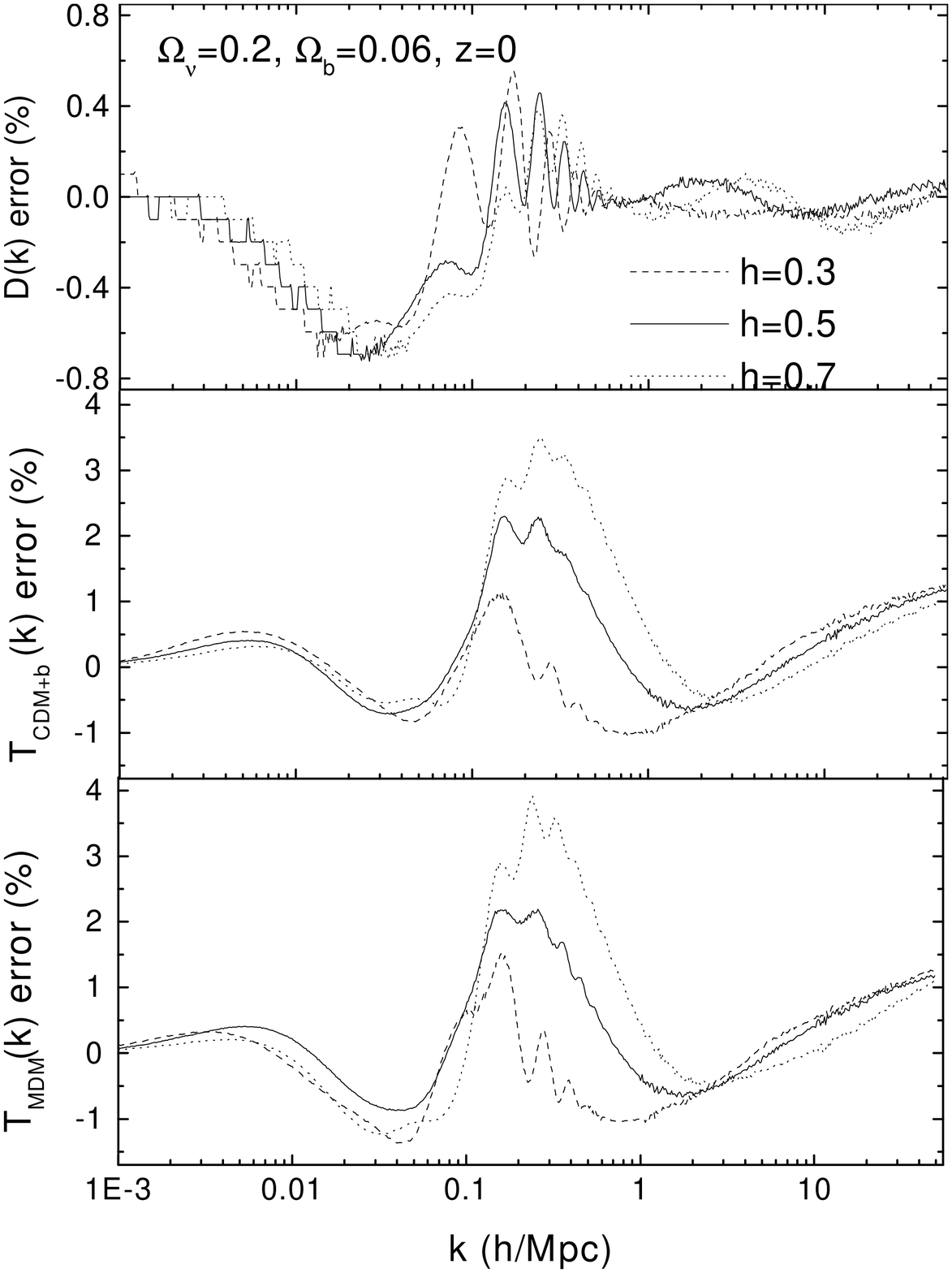}
\caption{Fractional residuals of the analytic approximation  of $D(k)$,
$T_{CDM+b}(k)$ and $T_{MDM}(k)$ 
transfer function for different values of the Hubble parameter, $h$.}
\end{figure}

A similar situation occurs also for the dependence of $T_{MDM}(k)$ 
on $h$. Since the $h$-dependence is included properly  in $k_F$ and 
$k_{nr}$, $D(k)$ does not require any correction in the asymptotic regimes. 
Only a tiny correction of $D(k)$ in the intermediate range,
 $k$ ($0.01<k<1$) is necessary to  minimize the residuals. 
By  numerical experiments  we find that this can be achieved by 
multiplying $\alpha_{1},...\alpha_{4}$
by the factors $C_i(h)$ which are approximated by second order polynomial
on $h/0.5$ with coefficients determined by $\chi^2$
minimization (see Appendix). The fractional residuals of
$D(k)$ for different $h$ are shown in Fig.~9 (top panel), they remain
stable in the range $0.3\le h\le 0.7$. But the fractional residuals of 
$T_{MDM}(k)$ slightly grow (about 2-3\%, bottom Fig.~9) in the range 
$0.1\le k \le 1$ when $h$ grows from 0.3 to 0.7. This is caused by 
the fractional residuals of  $T_{CDM+b}(k)$ (see middle panel).  
    
\subsection{Dependence on $N_{\nu}$.}

\begin{figure}[th]
\epsfxsize=8truecm
\epsfbox{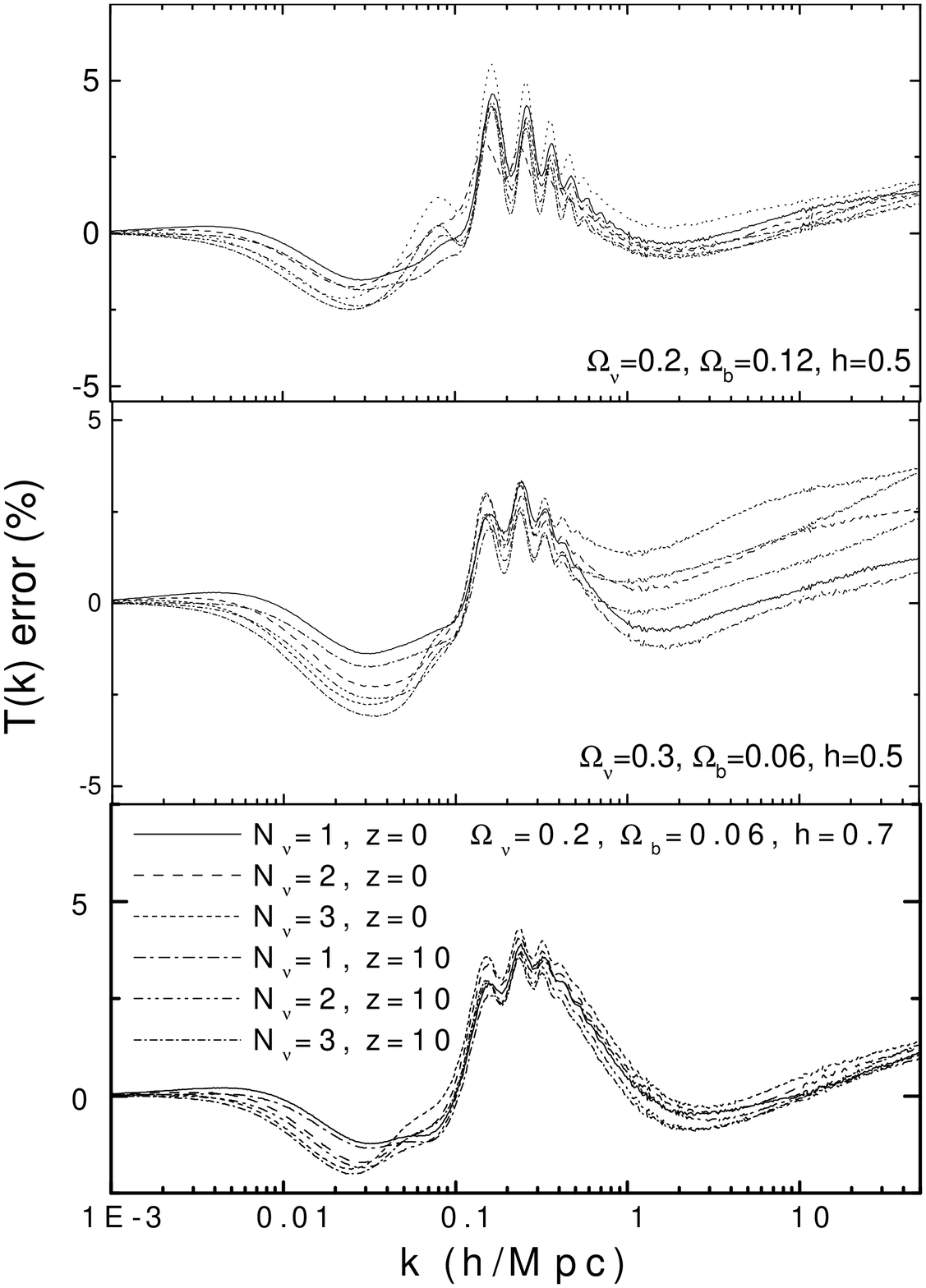}
\caption{
Fractional residuals for the analytic approximation  of the MDM
transfer function of density perturbations for different numbers  of
massive neutrino species, $N_{\nu}$, in models with different values
of $\Omega_{\nu}$, $\Omega_{b}$ and $h$.}
\end{figure}

The dependence of $D(k)$ on the number  of massive neutrino species, $N_{\nu}$, 
is taken into account in our analytic approximation by the
corresponding dependence of the  physical 
parameters $k_{nr}$ and $k_F$ (see Eq.(6)). It has the correct asymptotic
behaviour on small and large scales but rather large residuals in 
the intermediate region $0.01<k<10$.
 Therefore, the coefficients $\alpha_{i}$
($i=1,...,4$) must be corrected for $N_{\nu}$. To achieve this, 
we multiply each
$\alpha_{i}$ by a factor $D_{i}(N_{\nu})$ ($\sim 1$) which we
determine  by $\chi^2$ minimization. These factors  depend on
$N_{\nu}$ as second order polynomials. They are given in the Appendix. In
Fig.~10 we show the fractional residuals of $T_{MDM}(k)$ for different
numbers of massive neutrino  species, $N_{\nu}$, and several values of
the parameters $\Omega_{\nu}$, $\Omega_b$, $h$ and $z$.  The
performance for $N_{\nu}=2,3$ is approximately the same as for $N_{\nu}=1$.

\section{Performance} 
 
\begin{figure}[th] 
\epsfxsize=8truecm 
\epsfbox{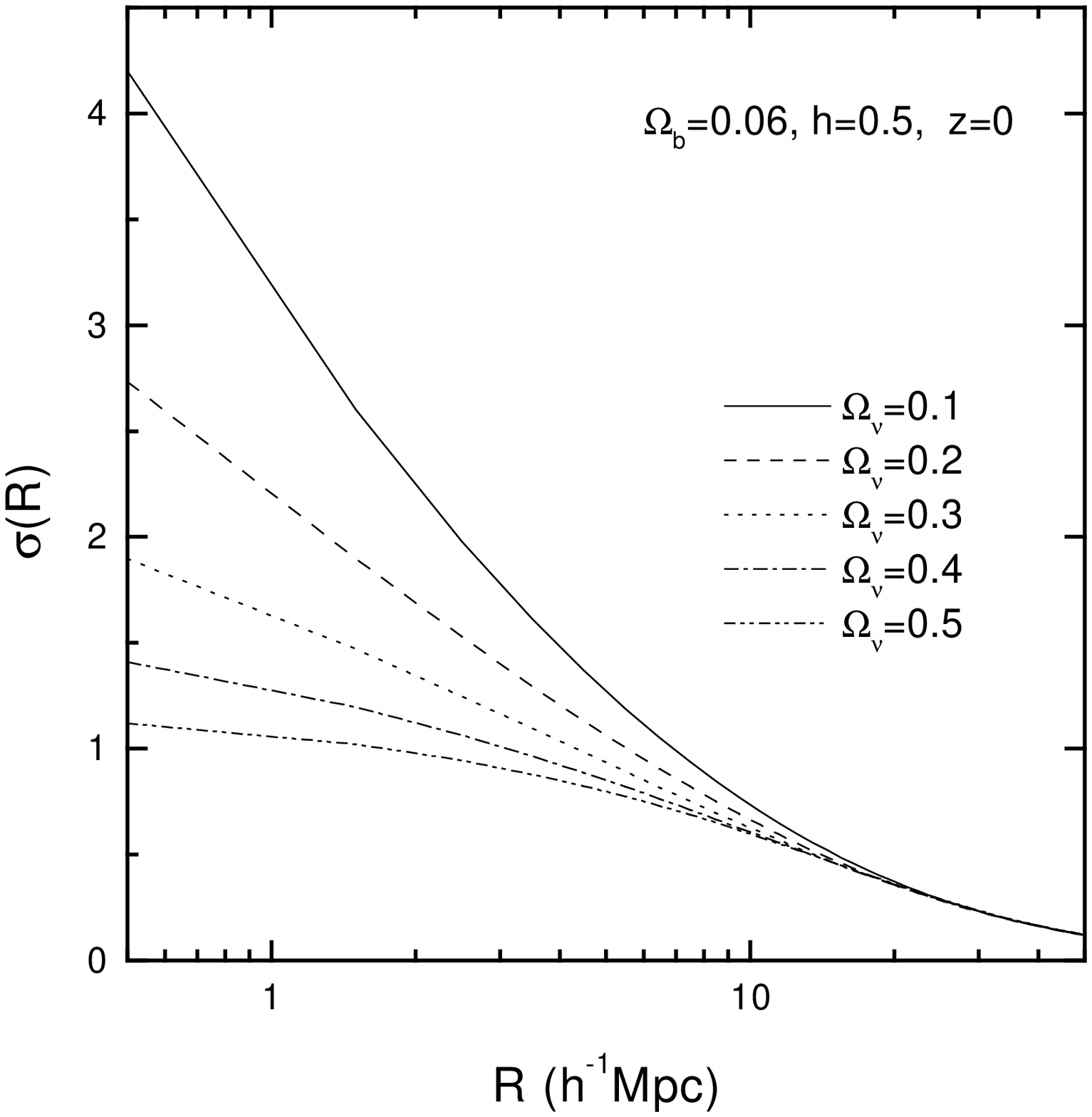} 
\caption{The variance of density perturbations 
smoothed by a top-hat filter of radius $R$ for different MDM models. 
Numerical transfer functions $T_{MDM}$ are shown as thick 
lines and the analytic approximations $T_{MDM}(k)=T_{CDM+b}D(k)$ are thin 
lines which are superimposed on the numerical results (The residuals
are not visible on this scale.). } 
\end{figure} 
 
\begin{figure}[th] 
\epsfxsize=8truecm 
\epsfbox{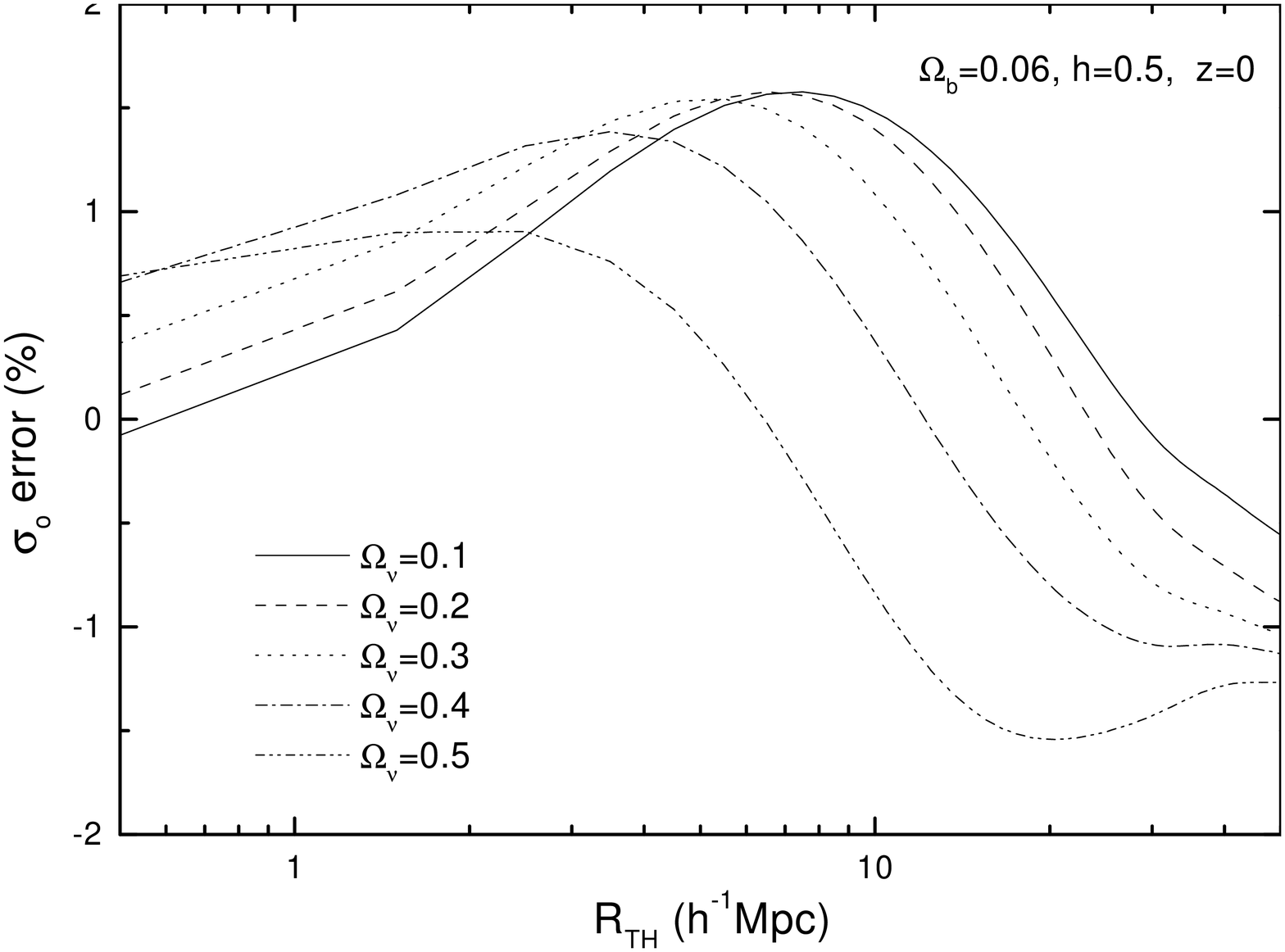} 
\caption{The fractional residuals of $\sigma(R)$ defined
by $(\sigma(R)-\sigma_{CMBfast}(R))/\sigma_{CMBfast}(R)$ 
for different values of $\Omega_{\nu}$ with all other parameters fixed.} 
\end{figure} 
 
The analytic approximation of $D(k)$ proposed here has maximal fractional 
residuals of less than $ 5\%$ in the range $0.01\le k \le 1$. It is
oscillating around the exact numerical result (see Fig.~4), which 
essentially reduces the fractional residuals of integral quantities
like $\sigma(R)$. Indeed, the mean square density perturbation 
smoothed by a top-hat filter of radius $R$ 
$$\sigma^{2}(R)={1\over 2\pi^{2}}\int_{0}^{\infty}k^{2}P(k)W^{2}(kR)dk,$$ 
where $W(x)=3(\sin x-x \cos x)/x^3$, $P(k)=AkT_{MDM}^{2}(k)$ (Fig.11) 
has fractional residuals which are only about  half the residuals
of the transfer function (Fig.12). 
To normalize  the power spectrum to the 4-year COBE data we have 
used the fitting formula by \cite{bun97}. 

The accuracy of  $\sigma(R)$ obtained by
our analytic approximation is better than $2\%$ for a wide range of 
$\Omega_{\nu}$ for $\Omega_{b}=0.06$ and $h=0.5$. Increasing  
$\Omega_{b}$ slightly degrades the approximation for $N_{\nu}> 1$, 
but even for a  baryon content as high as $\Omega_{b} \sim 0.2$,
 the fractional residuals of 
$\sigma(R)$ do not exceed $5\%$. Changing  $h$ in the range $0.3-0.7$ 
and $N_{\nu}=1-3$ do also not reduce the accuracy of $\sigma(R)$
beyond this limit. 
 
\begin{figure}[th] 
\epsfxsize=8truecm 
\epsfbox{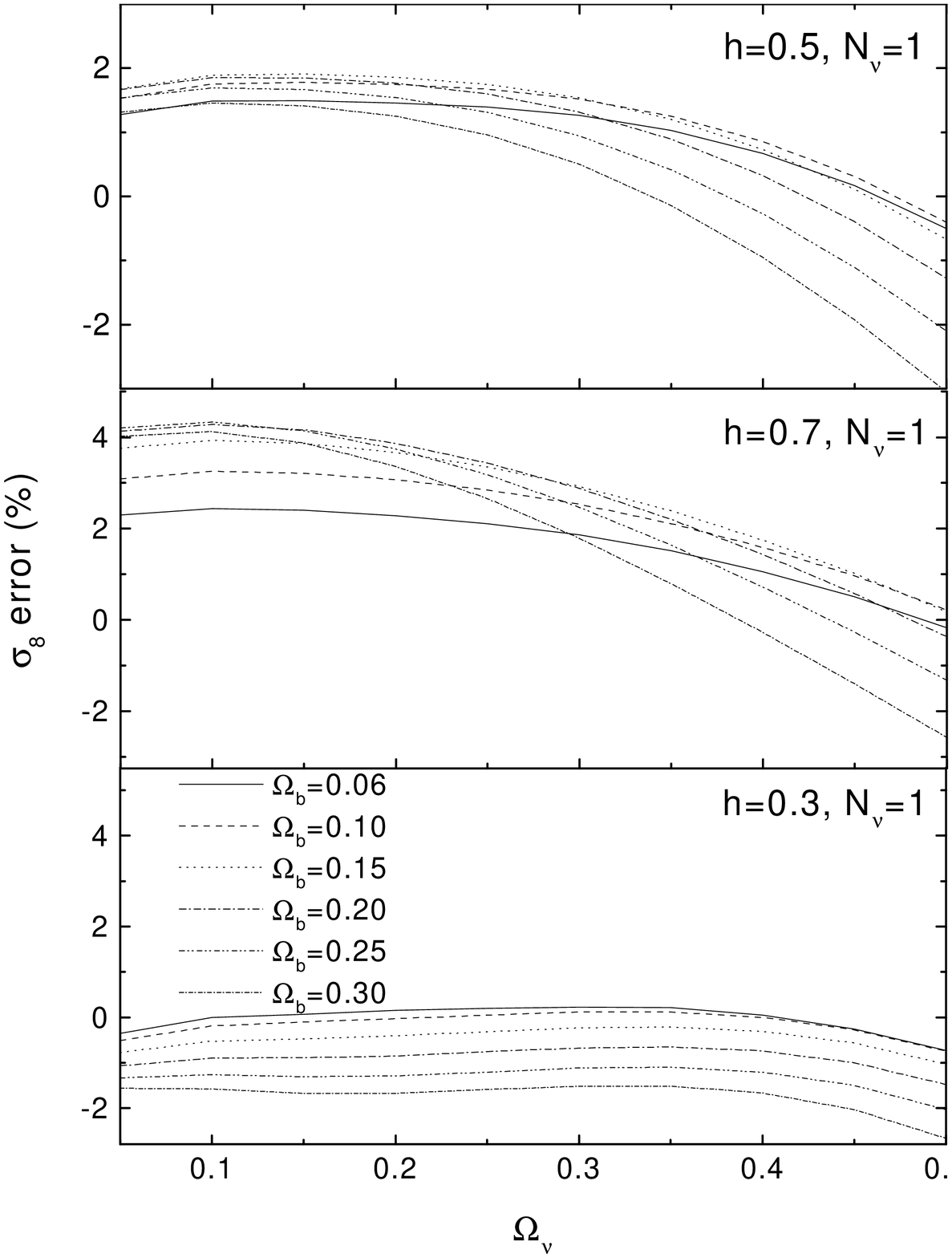} 
\caption{Deviation of $\sigma_{8}$ of our analytic
approximation for $T_{MDM}(k)$ from the numerical result
for different values of $\Omega_{\nu}$, $\Omega_{b}$ and $h$ ($N_{\nu}=1$). } 
\end{figure} 
 
\begin{figure}[th] 
\epsfxsize=8truecm 
\epsfbox{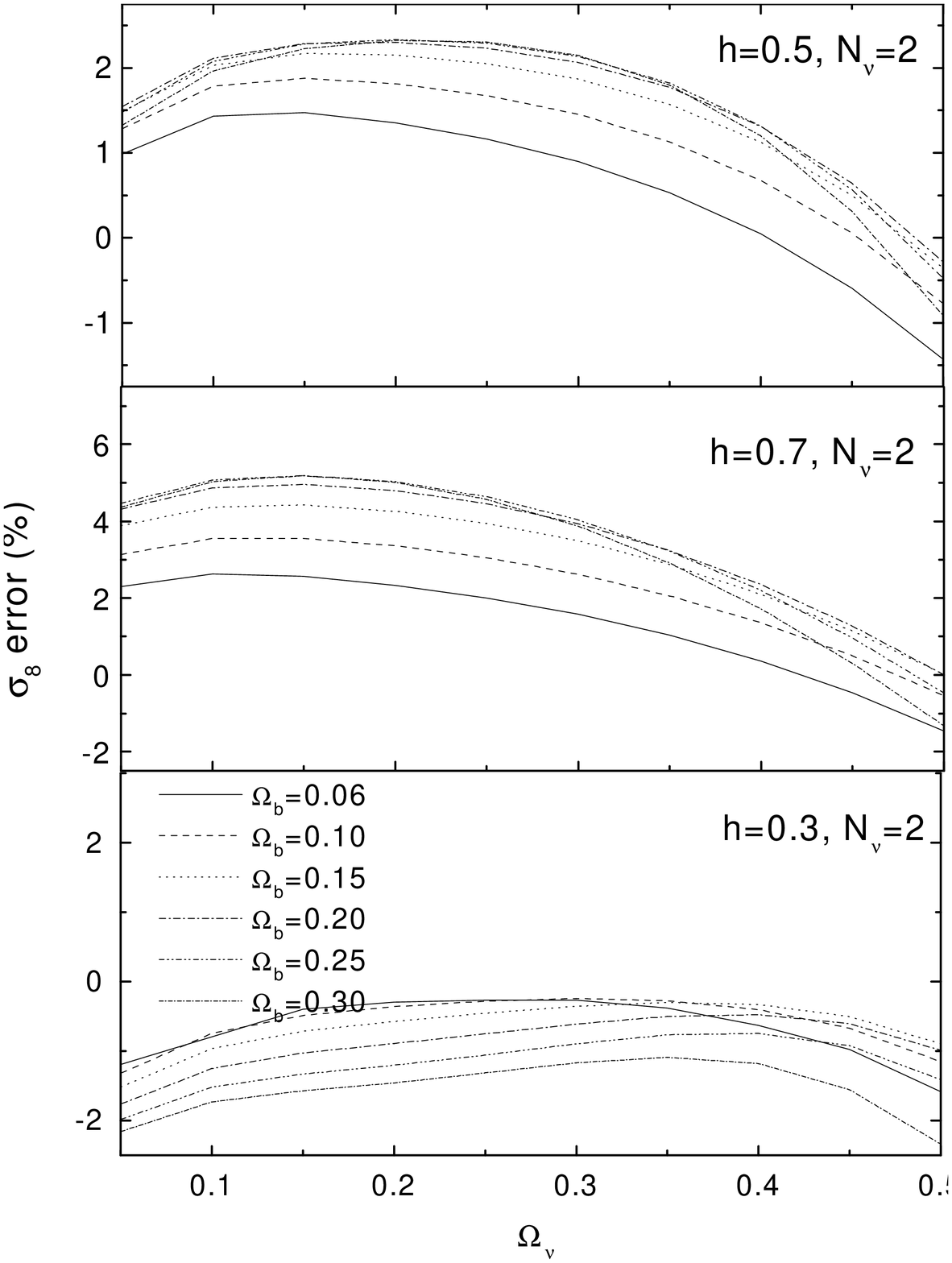} 
\caption{Same as Fig.13, but for $N_{\nu}=2$.} 
\end{figure} 
 
\begin{figure}[th] 
\epsfxsize=8truecm 
\epsfbox{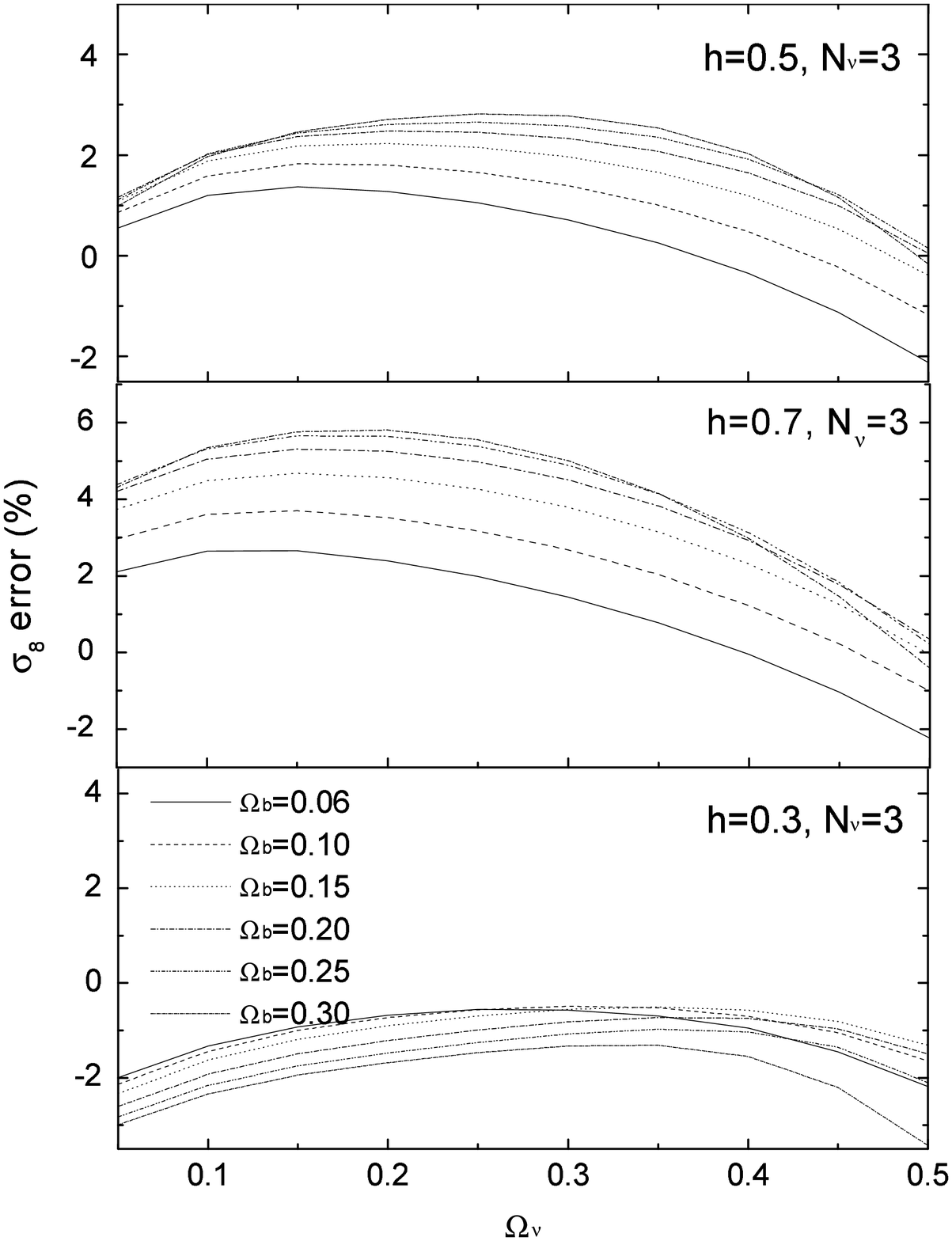} 
\caption{Same as Fig.13, but for $N_{\nu}=3$.} 
\end{figure}

We now evaluate the quality and fitness of our approximation in the four dimensional
space of parameters. 
We see in Fig.12 that the largest errors of our approximation  for 
$\sigma(R)$ come from scales of $\sim 5-10h^{-1}Mpc$\footnote{Actually, the
oscillations in the error of $T(k)$ are somewhat misleading: they
 are mainly due to
baryonic oscillations in the numerical $T(k)$ entering the denominator for
the error estimate, so that a slight shift of the phase enhances the
error artificially. This is why we concentrate on the error of 
$\sigma (R)$ (otherwise the error  estimate of T(k) should be 
averaged, see {\it e.g.} EH2).}. 
Since these scales are used for the evaluation of
the density perturbation amplitude on galaxy cluster scale, 
it is important to know how
accurately we reproduce them. The quantity $\sigma_{8}\equiv
\sigma(8h^{-1}Mpc)$ is actually  the most often used value to test
models.  We calculate it for the  set of 
parameters $0.05\le \Omega_{\nu}\le 0.5$, 
$0.06\le \Omega_{b} \le 0.3$, $0.3\le h \le 0.7$ and $N_{\nu}=1,\;2,\;3$ 
by means of our analytic approximation  and numerically. 
The relative deviations of $\sigma_{8}$ calculated with our $T_{MDM}(k)$ 
from the numerical results are shown in Fig.13-15. 
 
As one can see from Fig.~13, for $0.3\le h\le 0.7$ and 
$\Omega_{b}h^{2}\le 0.15$ the largest error in $\sigma_8$
for models with one sort of massive neutrinos $N_{\nu}=1$ does 
not exceed $4.5\%$ for  $\Omega_{\nu}\le 0.5$. 
Thus, for values of $h$ which are followed by direct measurements of the
Hubble constant, the range of $\Omega_{b}h^{2}$ where 
the analytic approximation is very accurate for $\Omega_{\nu}\le 0.5$ 
is six times as wide as the range given by nucleosynthesis constraints, 
($\Omega_{b}h^{2}\le 0.024$, \cite{tyt96}). This is important if one
wants to determine  cosmological parameters by the minimization 
of the difference between the observed and predicted characteristics of the 
large scale structure of the Universe.

For models with more than one species of massive neutrinos of equal mass
($N_{\nu}=2,3$), the accuracy of our analytic approximation is
slightly worse (Fig.~14,~15). But even for extremely high values of parameters 
$\Omega_{b}=0.3$, $h=0.7$, $N_{\nu}=3$ the error in
$\sigma_{8}$ does not exceed $6\%$.
 
In redshift space the accuracy of our analytic approximation is stable and 
quite high for redshifts of up to $z= 30$. 
 
\section{Comparison with other analytic approximations} 
 
\begin{figure}[th] 
\epsfxsize=8truecm 
\epsfbox{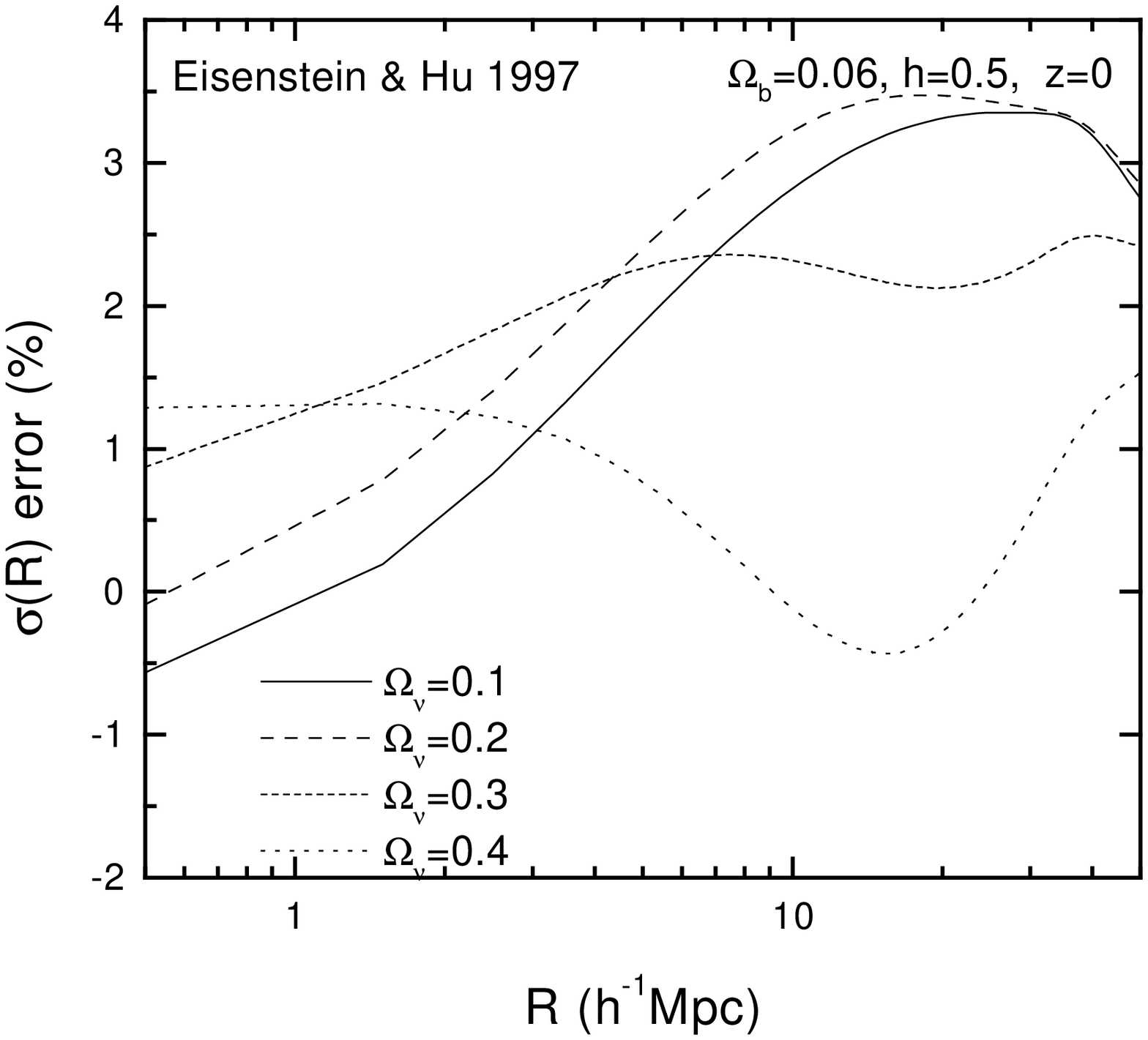} 
\caption{Fractional residuals of $\sigma(R)$ calculated by the
analytic approximation of EH2 for the same parameters 
as in Fig.~12.} 
\end{figure} 
 
\begin{figure}[th] 
\epsfxsize=8truecm 
\epsfbox{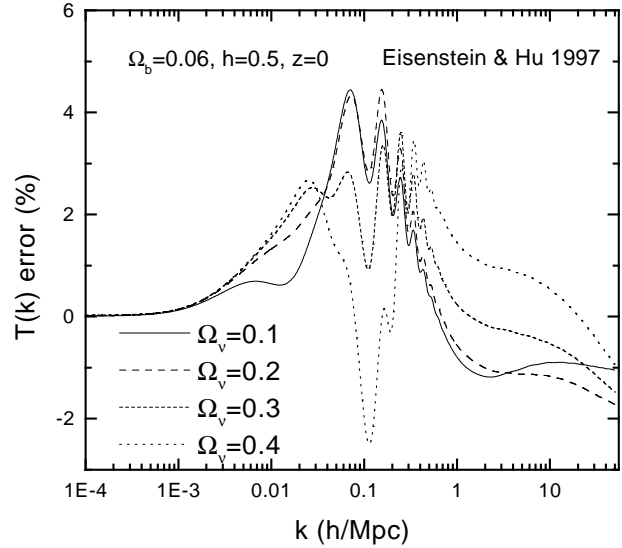} 
\caption{ 
Fractional residuals of the analytic approximation by EH2
of the MDM transfer function for the same parameters 
as in Fig.16. For comparison see Fig.6.} 
\end{figure} 
 
\begin{figure}[th] 
\epsfxsize=8truecm 
\epsfbox{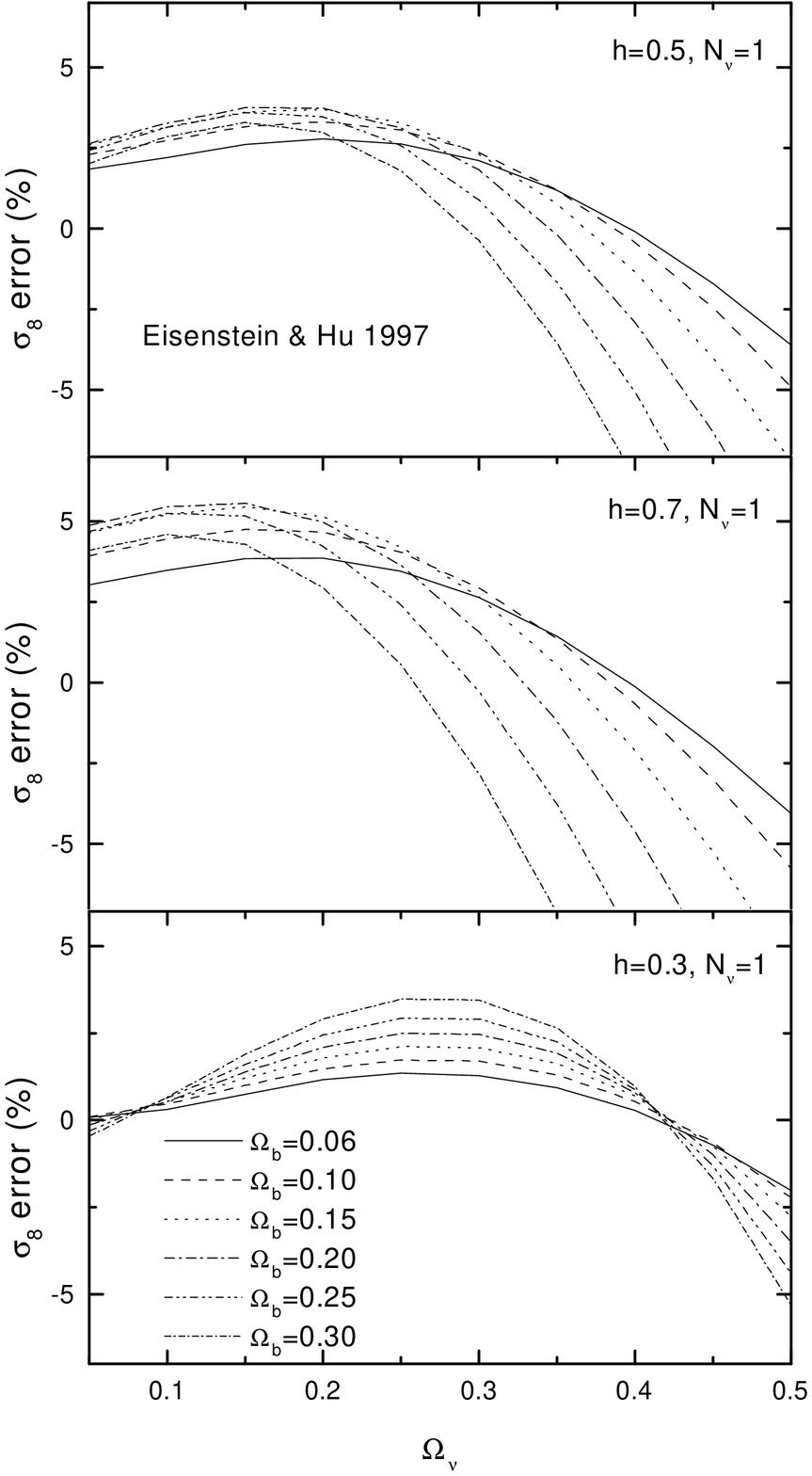} 
\caption{Deviation of $\sigma_{8}$ as obtained by the fitting formula
of EH2 from numerical results
for different values of $\Omega_{\nu}$, $\Omega_{b}$ and $h$ 
($N_{\nu}=1$). See Fig.~13 for comparison.} 
\end{figure} 
 
We now compare the accuracy of our analytic approximation with those 
cited in the introduction. For comparison with Fig.~12 the 
fractional residuals of $\sigma(R)$ calculated with the
analytic approximation of $T_{MDM}(k)$ by EH2
are presented in Fig.~16. Their approximation is only slightly less 
accurate ($\sim 3\%$) at scales 
$\ge 10 Mpc/h$. In Fig.~17 the fractional residuals of the 
EH2 approximation of $T_{MDM}(k)$ are 
shown for the same cosmological parameters as in Fig.~16. 
For $\Omega_{\nu}=0.5$ (which is not shown) the deviation from 
the numerical result is $\ge 50\%$ at $k\ge 1h^{-1}$Mpc, and the 
EH2 approximation completely
breaks down in this region of parameter space.
 
The analog to Fig.~13 ($\sigma_8$) for the fitting formula  of
EH2 is shown in Fig.~18 for different values of 
$\Omega_{b}$, $\Omega_{\nu}$ and $h$. 
 Our analytic approximation of $T_{MDM}(k)$ is
 more accurate than EH2
in the range $0.3\le \Omega_{\nu}\le 0.5$ for all $\Omega_{b}$ ($\le 0.3$). 
For $\Omega_{\nu}\le 0.3$ the accuracies of $\sigma_{8}$  
are comparable. 
 
\begin{figure}[th] 
\epsfxsize=8truecm 
\epsfbox{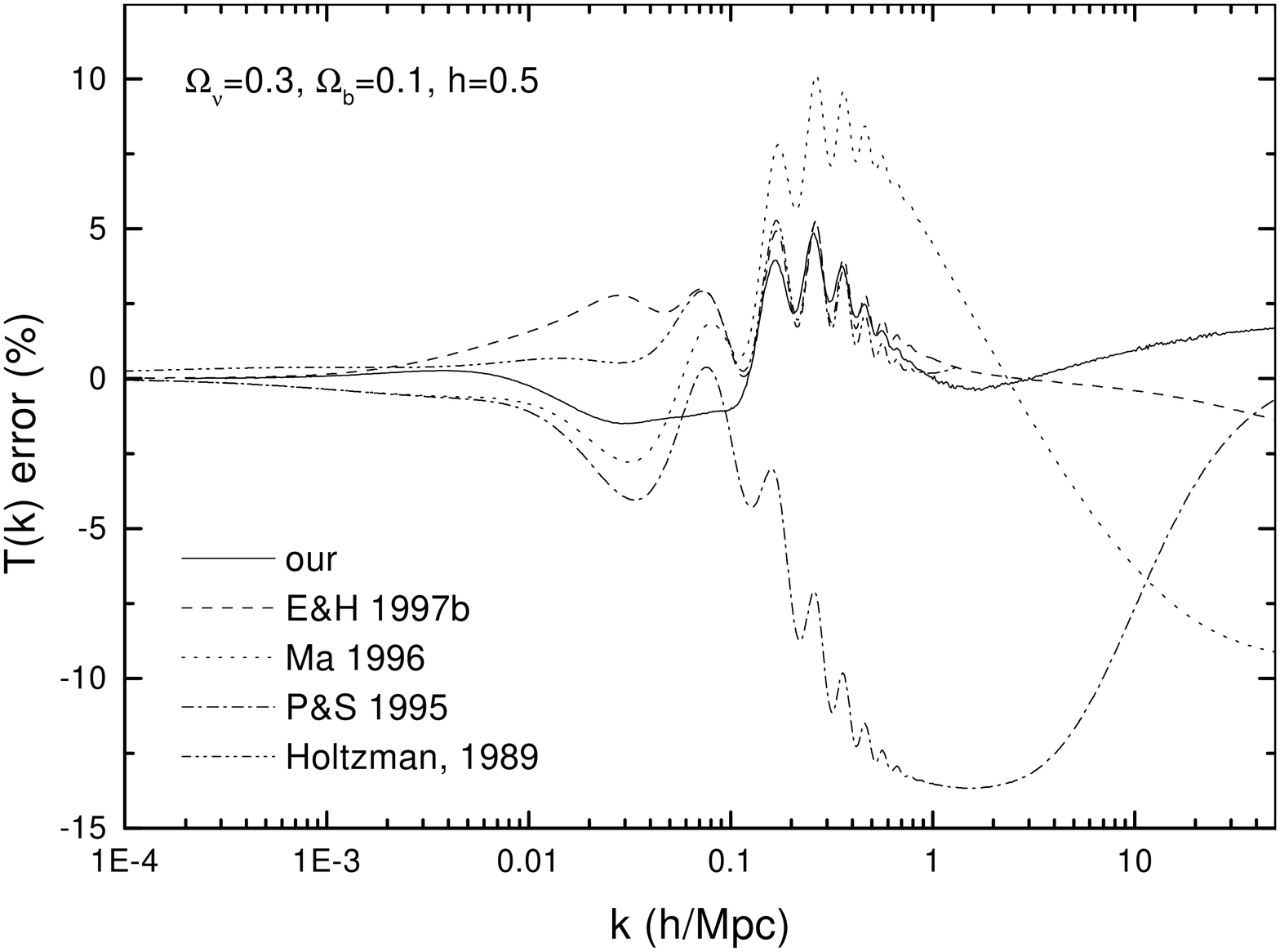} 
\caption{ 
Fractional residuals of different analytic approximations  for the
MDM transfer function at $z=0$ for one flavor of massive neutrinos.} 
\end{figure} 
 
\begin{figure}[th] 
\epsfxsize=8truecm 
\epsfbox{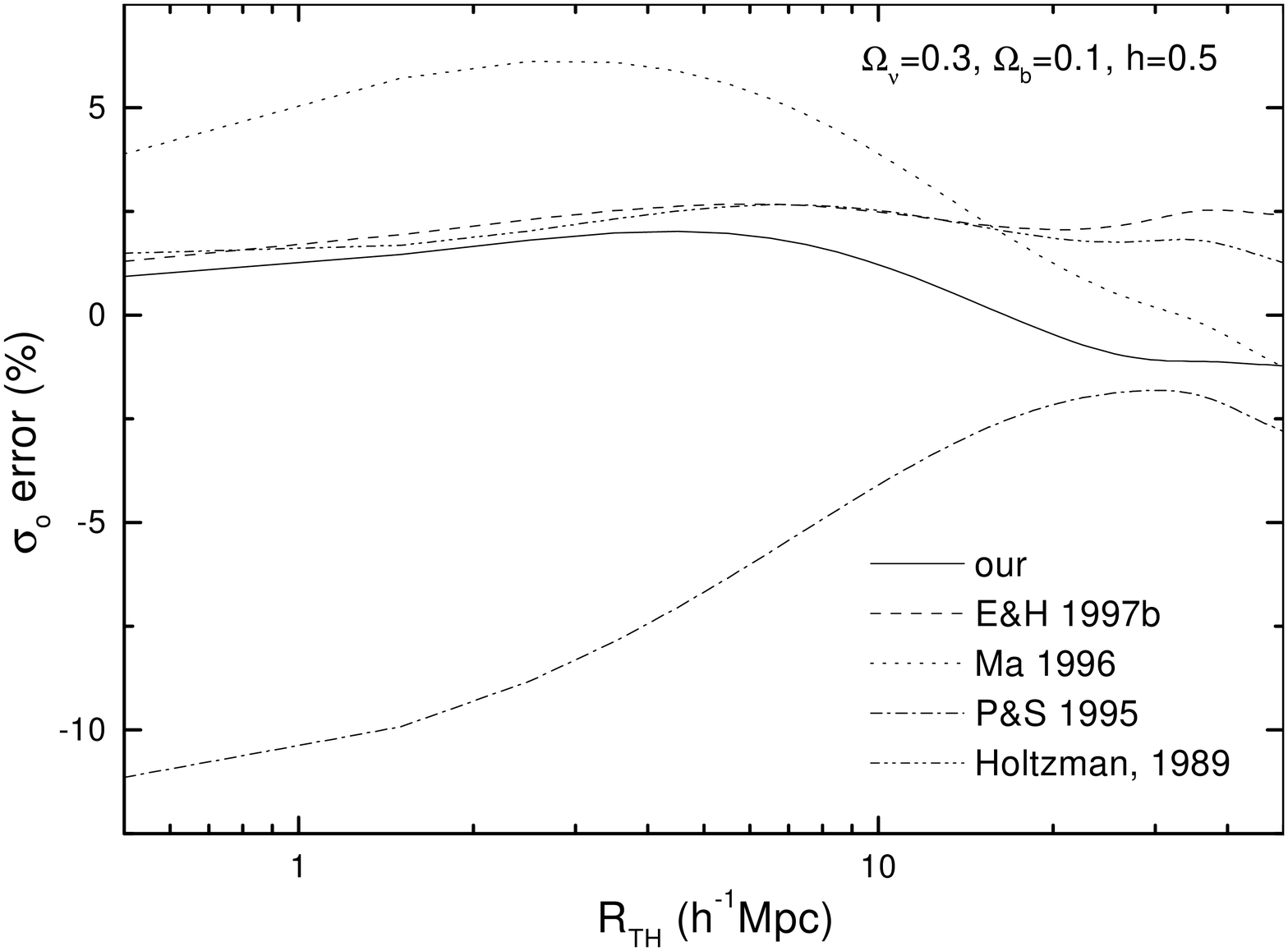} 
\caption{Fractional residuals of $\sigma(R)$ calculated with the  
same analytic approximations as Fig.~19.} 
\end{figure} 
 
To compare the accuracy of the analytic approximations for $T_{MDM}(k)$ 
given by \cite{hol89}, Pogosyan $\&$ Starobinsky 1995, \cite{ma96}, EH2 
with the one presented here, we
determine the transfer functions for the fixed set of  parameters 
($\Omega_{\nu}=0.3$, $\Omega_{b}=0.1$, $N_{\nu}=1$, $h=0.5$) 
for which all of them are reasonably accurate. Their deviations (in \%) from 
the numerical transfer function are shown in 
Fig.~19. The deviation of the variance of density fluctuations
for different smoothing scales from the numerical result 
is shown in Fig.~20. Clearly, our analytic approximation 
of $T_{MDM}(k)$ opens the possibility to determine the spectrum and
its  momenta more accurate in wider range of scales and parameters.

\section{Conclusions} 
 
We propose an analytic approximation for the linear  power spectrum of density 
perturbations in MDM models based on a correction of the approximation
by EH1 for CDM plus baryons. Our formula
is more accurate than previous ones (\cite{pog95,ma96}, EH2) for matter 
dominated Universes ($\Omega_{M}=1$) in a wide range of parameters: 
$0\le \Omega_{\nu}\le 0.5$, $0\le \Omega_{b}\le 0.3$, $0.3\le h\le
0.7$ and $N_{\nu}\le 3$. For models with one, two or three flavors of massive 
neutrinos ($N_{\nu}=1,\;2,\;3$) it is significantly more accurate than  the 
approximation by EH2 and has a relative error  
$\le 6\%$ in a wider range for $\Omega_{\nu}$ (see Figs.~13,~18).

The analytic formula given in this paper provides an essential tool
for testing a broad class of MDM models by comparison with
different observations like the galaxy power spectrum, cluster
abundances and evolution, clustering properties of Ly-$\alpha$ lines
etc. Results of such an analysis are presented elsewhere.

Our analytic approximation for  $T_{MDM}(k)$ is available in the form 
of a FORTRAN code and can be requested at 
{\bf novos@astro.franko.lviv.ua} or copied from 
{\bf http://mykonos.unige.ch/$\boldmath{\sim}$durrer/}

\medskip
 
{\it Acknowledgments} This work is part of a project supported by the
Swiss National Science Foundation (grant NSF 7IP050163). 
B.N. is also grateful to DAAD 
for financial support (Ref. 325) and AIP for hospitality,  where the 
bulk of the numerical calculations were performed. V.L. acknowledges a
partial support of the Russian Foundation for Basic Research (96-02-16689a).

\onecolumn{ 
\begin{appendix}{\bf Appendix} 
 
The best fit coefficients $a_i(\Omega_{\nu})$, $b_i(\Omega_{\nu})$, 
$c_i(\Omega_{\nu})$,  $B_i(\Omega_b)$, $C_{i}(h)$ and $D_i(N_\nu)$:

$$a_{1}=1.24198-3.88787\Omega_{\nu}+28.33592\Omega_{\nu}^2-70.9063\Omega_{\nu}^2
+84.15833\Omega_{\nu}^4-41.16667\Omega_{\nu}^5\;,$$ 
$$a_{2}=0.7295-3.6176\Omega_{\nu}+21.45834\Omega_{\nu}^2-54.63036\Omega_{\nu}^3+70.80274 
     \Omega_{\nu}^4-35.20905\Omega_{\nu}^5\;,$$ 
$$a_{3}=0.28283-0.53987\Omega_{\nu}+5.80084\Omega_{\nu}^2-14.18221\Omega_{\nu}^3
+16.85506\Omega_{\nu}^4-8.77643\Omega_{\nu}^5\;,$$ 
$$a_{4}=0.48431+1.89092\Omega_{\nu}-4.04224\Omega_{\nu}^2+8.09669\Omega_{\nu}^3
-10.05315\Omega_{\nu}^4+5.34405\Omega_{\nu}^5~.$$ 

\medskip 
 
$$b_{1}=0.2667-1.67\Omega_{\nu}+3.56\Omega_{\nu}^2-3.1\Omega_{\nu}^3\;,$$ 
$$c_{1}=0.008-0.055\Omega_{\nu}+0.135\Omega_{\nu}^2-0.124\Omega_{\nu}^3\;,$$ 

$$b_{2}=0.226-0.47\Omega_{\nu}+1.27\Omega_{\nu}^2-1.39\Omega_{\nu}^3\;,$$ 
$$c_{2}=0.004-0.026\Omega_{\nu}+0.053\Omega_{\nu}^2-0.039\Omega_{\nu}^3\;,$$ 
$$b_{3}=0.076-0.258\Omega_{\nu}+0.215\Omega_{\nu}2\;,$$ 
$$c_{3}=0.0026-0.0205\Omega_{\nu}+0.055\Omega_{\nu}^2-0.051\Omega_{\nu}^3\;,$$ 
$$b_{4}=0.0158-0.055\Omega_{\nu}+0.0228\Omega_{\nu}^2\;,$$
$$c_{4}=0.00094-0.0072\Omega_{\nu}+0.018\Omega_{\nu}^2-0.016\Omega_{\nu}^3~.$$ 

\medskip  

$$B_{1}(\Omega_b)= 1.202-0.2065(\Omega_{b}/0.06)+0.005(\Omega_{b}/0.06)^2\;,$$  
$$B_{2}(\Omega_b)=1.033-0.036(\Omega_{b}/0.06)+0.003(\Omega_{b}/0.06)^2\;,$$  
$$B_{3}(\Omega_b)=1.166-0.17(\Omega_{b}/0.06)+0.005(\Omega_{b}/0.06)^2\;,$$   
$$B_{4}(\Omega_b)=0.97985+0.01525(\Omega_{b}/0.06)+0.00626(\Omega_{b}/0.06)^2~.$$

\medskip 

$$C_{1}(h)=1.09-0.09(h/0.5)\;,$$ 
$$C_{2}(h)=1.65-0.88(h/0.5)+0.23(h/0.5)^2\;,$$ 
$$C_{3}(h)=1.055-0.055(h/0.5)\;,$$ 
$$C_{4}(h)=0.981+0.03(h/0.5)-0.012(h/0.5)^2~.$$ 

\medskip 

$$D_{1}(N_{\nu})=1.315-0.431N_{\nu}+0.116N_{\nu}^2\;,$$  
$$D_{2}(N_{\nu})=1.108-0.225N_{\nu}+0.117N_{\nu}^2\;,$$ 
$$D_{3}(N_{\nu})=1.316-0.432N_{\nu}+0.116N_{\nu}^2\;,$$ 
$$D_{4}(N_{\nu})=1.256-0.302N_{\nu}+0.046N_{\nu}^2~.$$

\end{appendix} 
} 
\clearpage 
\twocolumn

\end{document}